\documentclass{PoSa}
\usepackage{lineno}
\usepackage{mathtools}
\usepackage{amsmath,amstext}
\usepackage{booktabs}
\usepackage{sidecap}
\usepackage{wrapfig}
\usepackage{setspace}
\usepackage{booktabs}

\title{Search for correlations of high-energy neutrinos
and ultra-high energy cosmic rays}

\ShortTitle{Search for correlations of high-energy neutrinos
and UHECRs}

\author{
The IceCube Collaboration\footnote{For collaboration lists, see PoS(ICRC2019) 1177.}, The Pierre Auger Collaboration$^{*}$, The Telescope Array Collaboration$^{*}$, The ANTARES Collaboration$^{*}$\\

{\itshape \href{http://icecube.wisc.edu/collaboration/authors/icrc19_icecube}{http://icecube.wisc.edu/collaboration/authors/icrc19\_icecube}}\\
{\itshape \href{http://www.auger.org/archive/authors_icrc_2019.html}{http://www.auger.org/archive/authors\_icrc\_2019.html}}\\
{\itshape \href{http://www.telescopearray.org/images/ICRC2019/TAAuthor_ICRC2019.pdf}{http://www.telescopearray.org/images/ICRC2019/TAAuthor\_ICRC2019.pdf}}\\
{\itshape \href{http://antares.in2p3.fr/Collaboration/index2.html}{http://antares.in2p3.fr/Collaboration/index2.html}}\\
E-mail: \email{anastasia.barbano@icecube.wisc.edu,lisa.schumacher@icecube.wisc.edu}
}

\abstract{
The sources of ultra-high energy cosmic rays (UHECRs) are still one of the
main open questions in high-energy astrophysics. If UHECRs are accelerated
in astrophysical sources, they are expected to produce high-energy photons and
neutrinos due to the interaction with the surrounding astrophysical medium or ambient radiation. In
particular, neutrinos are powerful probes for the investigation of the region of
production and acceleration of UHECRs since they are not sensitive to magnetic
deflections nor to interactions with the interstellar medium.
The results of three different analyses that correlate the very high-energy neutrino 
candidates detected by IceCube and ANTARES and the highest-energy
cosmic rays measured by the Pierre Auger Observatory and the Telescope Array
will be discussed. The first two analyses use a sample of high-energy neutrinos
from IceCube and ANTARES 
selected to have a significant probability to be of astrophysical origin.
The first analysis cross-correlates the arrival directions of these selected neutrino events and UHECRs. The second one is a stacked likelihood 
analysis assuming as stacked sources the high-energy neutrino directions
and looking for excesses in the UHECR data set around the directions of the
neutrino candidates. The third analysis instead uses a larger sample of neutrinos selected to look for neutrino
point-like sources. It consists of a likelihood method that looks for excesses
in the neutrino point-source data set around the directions of the highest-energy
UHECRs.\\

\vspace{4mm}
{\bfseries Corresponding authors:}
\speaker{A. Barbano}$^{1}$\\
{$^{1}$ \itshape DPNC, University of Geneva}\\

}

\FullConference{36th International Cosmic Ray Conference -ICRC2019-\\
		July 24th - August 1st, 2019\\
		Madison, WI, U.S.A.}

\begin{document}

\section{Introduction}\label{sec:intro}
  \vspace{-6pt}
Galactic accelerators like supernova remnants are nowadays believed to be the most likely sources for cosmic rays (CRs) below $10^{15}$ eV~\cite{Amato:2014xua}.
On the other hand, ultra-high energy (above $10^{18}$ eV) cosmic rays (UHECRs) originate from some yet-unidentified extra-galactic sources, as indicated by the recent Pierre Auger Observatory measurement of the first statistically significant large-scale anisotropy anisotropy above 8 EeV~\cite{Aab:2017tyv}. The most promising sources, although contradictory in some aspects~\cite{ObservatoryMichaelUngerforthePierreAuger:2017fhr, 2009JCAP...11..009L, Kotera:2015pya, Murase:2005hy, 2011ARA&A..49..119K, Fang:2015xhg}, are active galactic nuclei, gamma ray bursts and magnetized and fast-spinning neutron
stars. UHECRs accelerated in astrophysical sources are expected to produce high-energy photons and neutrinos when interacting with the ambient matter and radiation. 
In particular, due to their tiny cross section and their insensitivity to (inter-)galactic magnetic fields, 
neutrinos constitute an excellent probe to investigate the origin of UHECRs. A multi-messenger approach might hence lead to a deep insight in the search of UHECR origin and their acceleration mechanisms.
In these proceedings, we present the results of three analyses searching for a common origin of UHECRs and high-energy neutrinos using data from IceCube Neutrino Observatory, the ANTARES Collaboration, the Pierre Auger Observatory and the Telescope Array (TA) Collaboration.
The analyses are (1) a cross-correlation analysis that scans angular distances between UHECRs and high-energy neutrinos, (2) a neutrino-stacking correlation analysis with UHECR directions and (3) a UHECR-stacking correlation analysis with neutrino directions. 

\vspace{-6pt}
\section{Observatories and data samples}\label{sec:neutrino}
\begin{wrapfigure}{r}{0.6\textwidth}
\begin{center}
\vspace{-30pt}
\includegraphics[width=0.6\textwidth]{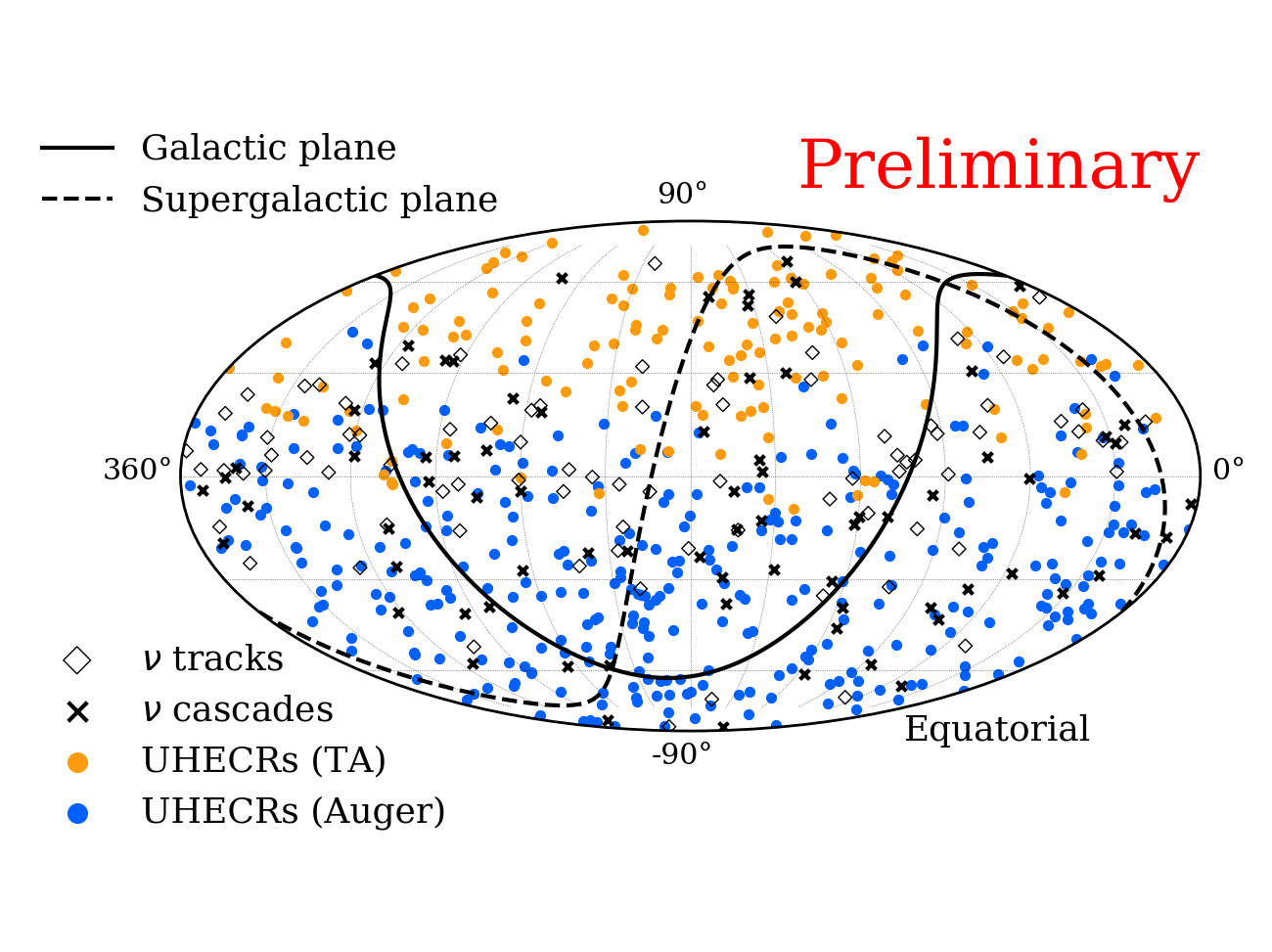}
\vspace{-40pt}
\caption{The UHECR events from TA and Auger are shown as orange and blue dots, respectively. The neutrino track- and cascade-like events from IceCube (HESE~\cite{wandkowsky_nancy_2018_1301088}, EHE~\cite{Aartsen:2018vtx}, 7-year through-going muons~\cite{Haack:2017dxi} samples) and ANTARES~\cite{Albert:2017ohr} are shown as black empty diamonds and crosses, respectively.}
\label{fig:eventSM}
\end{center}
\vspace{-15pt}
\end{wrapfigure}

  \vspace{-6pt}
{\bf IceCube}~\cite{Achterberg:2006md} is a 1-km$^{3}$ sized neutrino detector optimized for neutrino energies above $\sim$100 GeV,
located at the geographic South Pole at about 1.5 to 2.5 km deep in the glacial ice. It consists of
86 strings instrumented by 5160 photomultiplier tubes housed together with on-board digitization modules in pressure resistant
spheres. The first and the second analysis, which will be presented in the following, combine different IceCube datasets: 
(i) the 7.5-year data set of High-Energy Starting Events (HESE)~\cite{wandkowsky_nancy_2018_1301088}, (ii) the 9-year sample of Extremely High-Energy event alerts (EHE)~\cite{Aartsen:2018vtx}
and (iii) a complementary 7-year sample of through-going muons induced by charged-current interactions of $\nu_{\mu}$ candidates from the Northern sky~\cite{Haack:2017dxi}. The HESE sample is composed by 
76 shower-like events, characterized by an average angular resolution of $\sim$15$^{\circ}$ above 100 TeV,
and 26 track-like events, with an average angular resolution of $\sim$1$^{\circ}$~\cite{Kopper:2017zzm}. 
The IceCube realtime neutrino alert system is based on HESE and EHE selection by analyses looking for
cosmogenic neutrinos 
\cite{Aartsen:2016lmt}. 
The HESE alerts resulted in the evidence for the first neutrino emission in coincidence with a high-energy gamma-ray emission from the blazar named TXS 0506+056 \cite{IceCube:2018dnn}. The EHE analysis discovered the first observed PeV-scale neutrinos~\cite{Aartsen:2013bka}. The EHE alert event selection is sensitive to energies 
from about 500 TeV to 10 PeV and targets track-like events, which have good angular resolution 
($\leqslant$ 1$^{\circ}$). The used EHE sample is composed by 20 track-like events.
Finally, the through-going muons are composed by 35 tracks with $E \gtrsim $ 200 TeV, corresponding to 7 years of data from the 8-year sample presented in~\cite{Haack:2017dxi}.
Figure~\ref{fig:eventSM} shows the arrival directions of neutrino track- and cascade-like events described above, together with the ANTARES high-energy neutrinos and Auger and TA UHECR events described in the following. The third analysis uses (i) the 7-year neutrino point-source sample~\cite{Aartsen:2016oji} and (ii) the latest 3.5 years of the gamma-ray follow-up (GFU) sample~\cite{IceCube:2018cha}.
This combined track-like sample, selected for point-like source searches, consists of 1.4 million 
events recorded between 2008 and 2018. These are dominated in the Northern hemispheres by atmospheric $\nu_{\mu}$  and in the Southern hemisphere by atmospheric downgoing muons. The angular resolution is
$\textless 0.5^{\circ}$ above TeV energies~\cite{Aartsen:2016oji} . 

{\bf ANTARES}~\cite{Collaboration:2011nsa} is a neutrino telescope located in the Mediterranean Sea, composed by 
12 vertical strings anchored at the sea floor at a depth of $\sim 2400$ m, covering a total volume of \mbox{$\sim 0.03$ km$^3$}.
The strings are equipped with a total of 885 optical modules, each one housing a  
photomultiplier tube. The events used in analyses (1) and (2) are selected from the \mbox{9-year} point-source sample~\cite{Albert:2017ohr}, recorded between January 2007 and December 2015, while for analysis (3) they are selected from the 11-year point-source sample that includes events until 2017~\cite{Albert:2018kjg}. The samples include neutrino charged- and neutral-current interactions of all flavors. At energies of 10 TeV, the median
angular resolution for muon neutrinos is below 0.5$^{\circ}$. In particular, analyses ($1$) and ($2$) require an event
 signalness > 40\%, where the signalness is defined as the ratio of the number of expected astrophysical events
over the sum of the expected atmospheric and astrophysical events at a given energy proxy, where a spectrum $\phi = 1.01 (E/100\,{\rm TeV})^{-2.19} \cdot 10^{-18} {\rm GeV}^{-1} {\rm cm}^{-2} {\rm }^{-1} {\rm sr}^{-1}$ was used \cite{Aartsen:2017mau}.
This selection results in a total of three tracks and no cascades.

The {\bf Pierre Auger Observatory}~\cite{ThePierreAuger:2015rma} is located in Argentina at an average latitude of $\sim 35.2^{\circ}$ and a mean altitude of $\sim 1400$ m above the sea level.  
The Observatory is a hybrid detector combining the information from a large surface detector array (SD) and a fluorescence detector (FD). 
The SD array, spread over an area of 3000 km$^{2}$, is composed of 1660 water-Cherenkov detectors. The FD array consists of 27 telescopes at five peripheral buildings viewing the atmosphere over the SD array. 
The data sample used in this work consists of 324 events observed with the SDs from January 2004 to April 2017 with reconstructed energies > 52 EeV and zenith angle $\theta \leqslant 80^{\circ}$~\cite{PierreAuger:2014yba}, which translates into a field of view ranging from -90$^{\circ}$ to +45$^{\circ}$ in declination.  
At these energies the angular uncertainty is less than 0.9$^{\circ}$~\cite{2009NuPhS.190...20B}, the statistical uncertainty in the energy determination is better than 12$\%$~\cite{Abreu:2011pj} and the systematic uncertainty in the absolute energy scale is 14$\%$~\cite{Bruce:2019icrc}.

The {\bf Telescope Array} (TA) experiment~\cite{AbuZayyad:2012kk}, located 
in Utah (USA), detects cosmic rays with $E \textgreater 10^{18}$~eV. 
The surface array, composed by more than 500 scintillator detectors, extends over 
\mbox{700 km$^2$} of desert. In addition, there are three fluorescence telescope stations, 
instrumented with 12-14 telescopes each. The exposure of the detector covers the Northern Hemisphere
and the Southern Hemisphere up to -15$^{\circ}$. A total of 143 events with \mbox{energy $\geqslant$ 57 EeV} and \mbox{zenith angle $\leqslant$ 55$^{\circ}$}, recorded from May 2008 to May 2017, are 
used in this work~\cite{2013ApJ...768L...1A}. These events have about 1.5$^{\circ}$ angular resolution, $\sim$20\% energy resolution and a $\sim$22\% systematic uncertainty on the energy scale~\cite{2013ApJ...768L...1A}.

To account for the systematic energy shift in the absolute energy scale of UHECRs at the energies of interest in this work, in the likelihood analyses presented here, the Auger energy scale has been shifted by +14\% and the TA energy scale by -14\% following the latest studies of the Auger-TA joint working group~\cite{Biteau:2019aaf}.
\vspace{-10pt}
\section{Analysis methods}\label{sec:analysis}
\vspace{-6pt}
In the following, the three analyses are described separately. 
In general, shower- and track-like events are considered separately due to their 
different angular resolutions. Hence, separate $p$-values are provided by each of the analyses.
\vspace{-3pt}
\subsection{Cross-correlation analysis}\label{sec:xcorrAna}
The cross-correlation analysis counts the number, $n_{\text{obs}}$, 
of UHECR-neutrino pairs separated by less than an angular distance, $\Delta\alpha$. 
This number is compared to the simulated number, $n_{\text{exp}}$, of pairs within
the same $\Delta\alpha$ distance which are expected in the null-hypothesis scenario. 
Two separate null-hypotheses are investigated: (i) an isotropic distribution of UHECRs, 
obtained by generating isotropic CR datasets following the
exposure of the two experiments and (ii) an isotropic 
distribution of neutrinos, obtained by assigning randomly generated right-ascension values to the real neutrino events, hence preserving the 
declination-dependent acceptance of the neutrino observatories. 
The analysis is performed for different $\Delta\alpha$ values, from 1$^{\circ}$
to 30$^{\circ}$ in 1$^{\circ}$ steps.
The fraction of isotropic simulations with equal or larger number of pairs than in data gives a measurement of the probability (local $p$-value) that an observed excess of events arises by chance from an isotropic distribution. The final {\it p}-value of the most important excess is evaluated by accounting for the scan in angle. This analysis does not require any assumption on the (Galactic) magnetic field
(unlike the two analyses discussed in the following) 
since the scan on the angular distances between the neutrinos and the CRs 
already accounts for any possible angular separation due to magnetic deflections.

\vspace{-5pt}
\subsection{Neutrino-stacking correlation analysis with UHECR directions}\label{sec:NuStacking}

This analysis performs an unbinned-likelihood method by stacking
the arrival directions of the neutrinos and searching for
coincident sources of cosmic rays (CRs). The signal hypothesis assumes that \mbox{UHECRs} are correlated with high-energy neutrino directions. The background hypothesis is consistent with an isotropic distribution of UHECRs across the sky.
The logarithm of the likelihood 
function is defined as:
\begin{equation} \label{eq:LLHana2}
\text{ln} \,\mathcal{L}(n_s) = \sum_{i=1}^{N_{\text{Auger}}} \text{ln}{\bigg( }\frac{n_s}{N_{\text{CR}}}S^i_{\text{Auger}} + \frac{N_{\text{CR}}-n_s}{N_{\text{CR}}}B^i_{\text{Auger}}{\bigg) } +  \sum_{i=1}^{N_{\text{TA}}} \text{ln}{\bigg( }\frac{n_s}{N_{\text{CR}}}S^i_{\text{TA}} + \frac{N_{\text{CR}}-n_s}{N_{\text{CR}}}B^i_{\text{TA}}{\bigg) },
\end{equation}
where $n_s$ is the number of signal events, i.e. UHECRs correlated with neutrino directions, and is the only free parameter; $N_{\text{CR}} = N_{\text{Auger}} + N_{\text{TA}}$ is the total number of CR events. $S^i$ and $B^i$ are, respectively,
the signal and background probability distribution functions (PDFs) for each CR observatory.
The signal PDF, for the $i^{\text{th}}$ CR at a given direction $\vec{r}_i$ and with energy $E_i$, can
be expressed as:
\begin{equation} \label{eq:signalPDF}
S^i_{\text{CR observatory}} (\vec{r}_i,E_i) = R_{\text{CR observatory}}(\delta_i) \sum_{j=1}^{N_{\text{src}}} S_j(\vec{r}_i,\sigma(E_i)),
\end{equation}
where $R_{\text{CR observatory}}$ is the relative experiment exposure at a given event declination, $\delta_i$, 
$N_{\text{src}}$ is the number of stacked (neutrino) sources and $S_j(\vec{r}_i,\sigma(E_i))$ is the value of the normalized directional likelihood map for the $j^{\text{th}}$ source taken at position $\vec{r}_i$. The arrival direction of the $i^{\rm th}$ UHECR event is obtained by smearing the source
position with a two-dimensional Gaussian function with standard deviation $\sigma(E_i)$ calculated as $\sigma(E_i) = \sqrt{\sigma^2_{\text{CR observatory}}+\sigma^2_{\text{MD}}}$,
where $\sigma_{\text{CR observatory}}$ is the angular resolution of the CR observatory  
(0.9$^{\circ}$ for Auger and 1.5$^{\circ}$ for TA) and $\sigma_{\text{MD}} = D \times 100$ EeV/$E_{\text{CR}}$ is the energy-dependent Galactic magnetic deflection.
In the analysis, $D$ is assigned three benchmark values. In order to account for the differences of the 
Galactic magnetic field in the Northern and Southern hemispheres, two average deflection values 
are calculated, whereas previous analyses used all-sky average deflection values~\cite{Aartsen:2015dml,AlSamarai:2017pos}, by considering the Galactic magnetic field models of Pshirkov et al.~\cite{Pshirkov:2011um} and Jansson and
Farrar~\cite{Jansson:2012pc}. Assuming a pure proton-like CR sample with $E_{\text{CR}} = 100$ EeV, mean angular deflection values of 2.4$^{\circ}$ and 3.7$^{\circ}$ are obtained, for the North and South respectively. 
To account for possible heavier compositions of the CRs or larger contributions of the magnetic fields, the average deflection values in the North and in the South at $E_{\text{CR}} = 100$ EeV are increased by factors 2 and 3.
Finally, the background PDFs, $B_{\text{Auger}}$ and $B_{\text{TA}}$ in Eq.~\ref{eq:LLHana2}, represent 
the probability of observing a cosmic ray from a given direction assuming an isotropic flux. 
Therefore they are calculated from the Auger and TA normalized exposures. The test statistic (TS) is defined as TS = 2 ln $(\mathcal{L}(\hat{n}_s)/\mathcal{L}(\hat{n}_s = 0))$, where $\hat{n}_s$ denotes the optimized parameter.
\vspace{-3pt}

\subsection{UHECR-stacking correlation analysis with neutrino directions}
\label{sec:CRStacking} 

This analysis is based on an unbinned likelihood method for searching point-like neutrino sources~\cite{Aartsen:2016oji}, with additional information from stacking UHECR arrival directions. 
The neutrino events are weighted according to the relative experiment exposure~\cite{Adrian-Martinez:2015ver}.
The signal hypothesis assumes point-like neutrino sources to be spatially correlated with UHECR arrival directions, which are subject to a specific magnetic deflection hypothesis. 
The background hypothesis assumes that neutrino events are uniformly distributed over the whole sky.
The free signal parameters of the likelihood function, $\mathcal{L}$, are the numbers of neutrino signal events, $n_s$, and the spectral indices, $\gamma_s$, for each possible neutrino source at positions $\overrightarrow{x_s}$.
The logarithm of the likelihood function is defined as:
\begin{equation}
    \label{eq:anaC-llh}
    \ln \mathcal{L} 
    = \underbrace{\sum_{s=1}^{N_{\rm CR}}}_{\substack{\text{stacking,} \\ \text{step 3}}} \Bigg[ \underbrace{ \bigg(\sum_{i=1}^{N_\nu} 
    \ln \big( \frac{n_{s}}{N_\nu} S_i (\gamma_{s}, \overrightarrow{x_{s}} )       
  + \left(1-\frac{n_{s}}{N_\nu} \right )B_i (\overrightarrow{x_{s}}) \big)\bigg)}_{\substack{\text{neutrino data,} \\ \text{step 1}}}
  -  \underbrace{\frac{\left(\overrightarrow{x_{s}} - \overrightarrow{x}_{{\rm CR},s} \right)^2}{2\sigma(E_{{\rm CR},s})^2}}_{\substack{\text{UHECR data,} \\ \text{step 2}}} \Bigg] .
\end{equation}
The first part of the likelihood formula (step 1 in eq.~\ref{eq:anaC-llh}) is determined by information from neutrino data, where the sum runs over all experimentally measured neutrino candidates $N_\nu$.
The signal PDF, $ S_i $, describes a point-like neutrino source at position $\overrightarrow{x_{s}}$ with $n_s$ events following a certain spectral index, $\gamma_{s}$.
The index $s$ denotes one neutrino source as counterpart to one UHECR event, as described later.
The background PDF, $B_i$, is determined from experimental neutrino events whose right ascension coordinates were assigned random values.
Any information from UHECR data is contained in the spatial prior functions shown in the second part of the likelihood (step 2). 
In the first step,  the signal parameters $(n_{s}, \gamma_{s})$ are optimized \textit{without the spatial prior function} on grid positions 
covering the whole sky, based on IceCube's standard procedure for searching for point-like sources~\cite{Aartsen:2016oji}.
The result is translated into a TS map of the neutrino sky. The TS is defined as  
$\text{TS}(\overrightarrow{x_s}) = 2\ln \left(\mathcal L_\text{step 1}\left ( \hat{n}_s, \hat{\gamma}_s\right )/\mathcal L_\text{step 1} \left (n_s=0 \right ) \right) $, where $ ( \hat{n}_s, \hat{\gamma}_s )$ denote the optimized parameters. 
In a second step, the arrival direction of one UHECR event, $ \overrightarrow{x}_{{\rm CR},s}$, and the corresponding smearing, $\sigma(E_{{\rm CR},s})$ 
(cf. sec.~\ref{sec:NuStacking}) are used to construct a 2D-Gaussian function, which is logarithmically added to the TS map.
This results in an effective selection of the neutrino sky where the largest remaining TS spot is 
the most likely neutrino source counterpart to the selected UHECR event.
This is equivalent to optimizing $n_s$, $\gamma_s$ and the source position in presence of a spatial Gaussian prior function.
The third step is to repeat the procedure for all selected UHECRs, and the resulting TS values are summed to yield the final TS. 
This is equivalent to a stacking of independent neutrino sources selected by UHECR and relative deflection information.
Note that this procedure ensures that each UHECR event has one neutrino-source counterpart in its vicinity,
while allowing one neutrino source to be counterpart to several UHECRs in its vicinity.
Three different lower energy cuts $E_{\rm CR}\geq[70, 85, 100]~\mathrm{EeV}$ are applied to the combined UHECR sample in order to study a potential energy dependence in the final TS.
Magnetic deflection values of $D=3^\circ$ and $6^\circ$ are used uniformly over the whole sky.
\vspace{-2pt}
\vspace{-6pt}
\section{Results}\label{sec:results}
\vspace{-6pt}
\subsection{Cross-correlation analysis}
\begin{figure}[h!]
\vspace{-9pt}
\begin{center}
\includegraphics[width=0.49\textwidth]{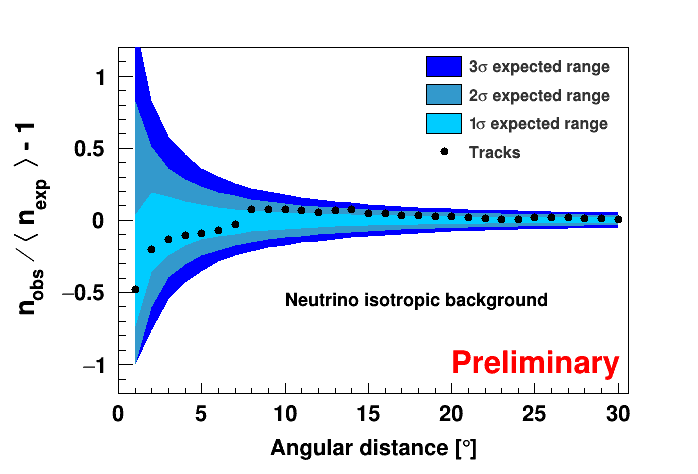}
\includegraphics[width=0.49\textwidth]{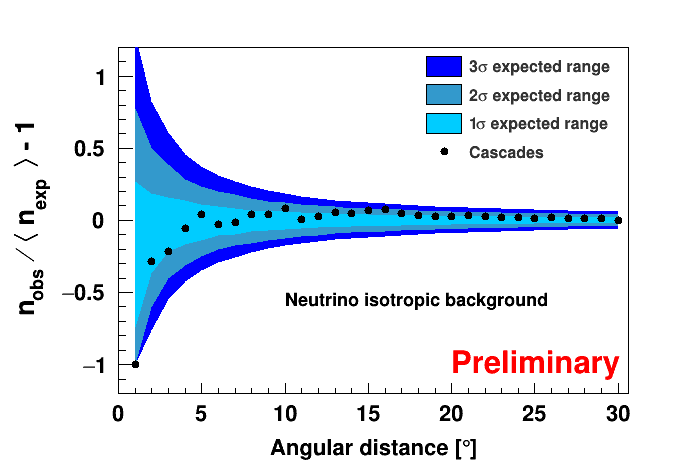}
\caption{Relative excess of pairs, $n_{\rm obs}/\langle n_{\rm exp}\rangle -1$, as a function of the maximum angular
separation between the neutrino and UHECR pairs, for track- (left) cascade-like (right) events in the case of an isotropic distribution of neutrinos. The different color bands stand for the regions containing the 1, 2 and 3$\sigma$ fluctuations from an isotropic distribution.} 
\label{fig:ana1}
\end{center}
\vspace{-10pt}
\end{figure}

The results of the scan in angle are shown in Fig.~\ref{fig:ana1}, where the relative number of observed pairs with respect to the expected value from an isotropic distribution of neutrinos is shown for tracks (left) and cascades (right). The maximum departure from the isotropy is found at 14$^{\circ}$ for tracks, where 582 pairs are observed, and at 16$^{\circ}$ for cascades, with observed 763 pairs. The post-trial {\it p}-values are 0.23 for tracks and 0.15 for cascades. The maximum departure from isotropy with respect to an isotropic distribution of UHECRs is found at 10$^{\circ}$ for tracks, where 303 pairs are observed and at 16$^{\circ}$ for cascades, with observed 763 pairs. The post-trial {\it p}-values are 0.84 for tracks and 0.18 for cascades. 
\subsection{Neutrino-stacking correlation analysis with UHECR directions}
The results are shown in Tab.~\ref{tab:ana2-res}. The most significant deviation
from an isotropic flux of CRs occurs for the magnetic deflection parameter set of D = (7.2$^{\circ}$, 11.1$^{\circ}$) with the high-energy cascade events. 
\begin{table}[!h]
\centering
\begin{tabular}{lrrrrrr}
\toprule
$D$     & (2.4$^{\circ}$, 3.7$^{\circ}$) & (4.8$^{\circ}$,7.4$^{\circ}$) & (7.2$^{\circ}$,11.1$^{\circ}$)\\
\midrule
tracks & underfluctuation & underfluctuation & underfluctuation \\
cascades & underfluctuation & 0.41 & 0.29 \\
\bottomrule
\end{tabular}
\caption{Pre-trial $p$-values for the neutrino-stacking analysis with the samples of high-energy tracks and cascades
assuming an isotropic flux of UHECRs.}
\label{tab:ana2-res}
\end{table}
The observed pre-trial $p$-value of 0.29 corresponds to 0.90 post-trial, obtained by considering multiple realizations of pseudo-experiments of randomly distributed CRs with \mbox{D = (7.2$^{\circ}$, 11.1$^{\circ}$)} and by accounting
for the trial factor due to having tested three different sets of magnetic deflections. Given these numbers, no sign of correlations in the arrival directions of UHECRs and neutrinos is found.
\vspace{-6pt}
\subsection{UHECR-stacking correlation analysis with neutrino directions}
Six $p$-values for each of the signal hypotheses described in sec.~\ref{sec:CRStacking} are calculated with respect to an isotropic neutrino flux, summarized in Tab.~\ref{tab:ana3-res}. 
All six $p$-values are well-compatible with the background expectation. The smallest $p$-value ($6\%$ for $D=6^\circ$ and $E_{\rm CR}\geqslant85$ EeV), after correction for the six correlated tests, becomes $16\%$.
\begin{table}[!h]
\centering
\begin{tabular}{lrrrrrr}
\toprule
$D~[{}^\circ]$        & \multicolumn{3}{c}{3} & \multicolumn{3}{c}{6} \\
$E_{\rm CR}$~[EeV] $\geqslant$ &  70  &  85  &  100 &  70  &  85  &  100 \\
\midrule
$p$-value               & 0.27 &  0.46 &  0.84 &  0.10 &  0.06 &  0.39 \\
\bottomrule
\end{tabular}
\caption{All pre-trial $p$-values for different UHECR energy cuts and deflection hypotheses.}
\label{tab:ana3-res}
\end{table}
\vspace{-8pt}

\vspace{-10pt}
\section{Discussion}\label{sec:discussion}
\vspace{-6pt}
The results of the three analyses presented here do not allow to conclude about the presence of possible correlations between arrival directions of UHECRs and high-energy neutrinos.
Previous analyses on reduced data sets had shown a post-trial $p$-value of 5.0 $\times$ 10$^{-4}$ for the cascades with a cross-correlation analysis, under the assumption of an isotropic flux
of UHECRs, and of \mbox{8.0 $\times$ 10$^{-4}$} for cascades with the neutrino-stacking analysis under a deflection
hypothesis of \mbox{D = 6$^{\circ}$}~\cite{Aartsen:2015dml}.
The absence of correlation found with the current data samples and discussed analysis hypotheses must be carefully interpreted, since it does not imply
an absolute lack of correlation in the origin of the two messengers.
The main uncertainties in the current analyses are the poor knowledge
of the Galactic magnetic field and the not yet conclusive understanding of the CR composition.
Furthermore, due to the GZK horizon, the largest distances covered by UHECRs are not expected to exceed 10-100 Mpc, depending on the CR composition. On the other hand, neutrinos can reach us from cosmological distances.
Finally, neutrinos originating at the cosmic rays acceleration sites are expected to carry few percents of the energy
of the original cosmic ray. Thus, the neutrinos observed
by IceCube and ANTARES might have been produced by
cosmic rays of much lower energy than the ones in the
datasets by Auger and TA.
For these reasons, only a few percent of neutrinos
may be expected to originate from the same astrophysical sources of the detected UHECRs.

\bibliographystyle{ICRC}
\begingroup
\setstretch{0.}
\bibliography{references}
\endgroup

\end{document}